\documentclass[sigconf, authorversion=true]{acmart}
\usepackage[english]{babel}
\usepackage[utf8]{inputenc}
\usepackage[T1]{fontenc}
\usepackage{setspace}
\usepackage[flushleft]{threeparttable} 
\usepackage{tabulary} 
\usepackage{dblfloatfix}
\usepackage{framed} 
\usepackage{xcolor} 

\usepackage{csvsimple}

\usepackage{graphicx}

\usepackage{hyperref} 
\hypersetup{
    pdfauthor={Christian Kröher, Sascha El-Sharkawy, and Klaus Schmid},
    pdftitle={KernelHaven - An Open Infrastructure for Supporting Software Product Line Engineering},
    pdfsubject={22st International Conference on Software Product Line},
    pdfkeywords={Software product lines; variability modeling; reverse engineering; static analysis},
    pdfpagemode={UseOutlines},
    pdfproducer={LaTeX},
    colorlinks=true,
    linkcolor=black,          
    citecolor=black,          
    filecolor=black,          
    urlcolor=black            
}

\usepackage{enumitem}
\setlist[itemize]{leftmargin=1em}
\setlist[enumerate]{leftmargin=3ex}
\setlist[description]{leftmargin=1em}

\colorlet{shadecolor}{gray!25}
\newcommand{\Command}[1]{
    \begin{snugshade*}
      \noindent #1
    \end{snugshade*}
}

\newcommand{\trim}{\looseness-1}
\newcommand{\ifDef}{\texttt{\#ifdef}}

\definecolor{cEclipseRed}{rgb}{0.6,0,0}
\definecolor{cEclipseGreen}{rgb}{0.25,0.5,0.35}
\definecolor{cEclipsePurple}{rgb}{0.5,0,0.35}
\definecolor{cEclipseBlue}{rgb}{0.25,0.35,0.75}

\usepackage{listings}
\lstdefinelanguage{j} {
    language=Java,
    keywordstyle=[2]{\color{gray!70!black}},
    morekeywords=[2]{@Override},
}

\lstdefinelanguage{properties} {
    tabsize=1,
    commentstyle=\color{blue}, 
    morecomment=[s][\color{cEclipsePurple}]{<}{>},
		morecomment=[l]{\	},
    morecomment=[l][\color{cEclipseGreen}]{\#},		
}

\usepackage{ifthen}
\newboolean{showToDos}
\setboolean{showToDos}{false} 
\ifshowToDos
  \usepackage{marginnote} 
  \let\marginpar\marginnote
  
  \newcommand{\ToDo}[1]{\noindent\fcolorbox{black}{yellow!20!white}{\parbox{.975\columnwidth}{#1}}}
  \newcommand{\SubjectedToDo}[2]{\ToDo{\textbf{#1:}#2}}
  
  \newcommand{\Outline}[1]{\color{blue}{#1}\color{black}}
  \newcommand{\Problem}[1]{\color{red}{#1}\color{black}}
  \newcommand{\secSize}[1]{[#1]}
  
  \usepackage[textsize=small,colorinlistoftodos]{todonotes}
  \newcommand{\userComment}[2]{\todo[color=#1!20,linecolor=#1!60,bordercolor=#1!60]{#2}}
  \newcommand{\ks}[1]{\userComment{orange}{\textbf{KS:} #1}}
  \newcommand{\se}[1]{\userComment{red}{\textbf{SE:} #1}}
  \newcommand{\ak}[1]{\userComment{blue}{\textbf{AK:} #1}}
  \newcommand{\ck}[1]{\userComment{green}{\textbf{CK:} #1}}

  \usepackage{pifont}
  
\else
  \newcommand{\ToDo}[1]{}
  \newcommand{\SubjectedToDo}[2]{}
  \newcommand{\Outline}[1]{}
  \newcommand{\Problem}[1]{}
  \newcommand{\secSize}[1]{}
  
  \newcommand{\ks}[1]{}
  \newcommand{\se}[1]{}
  \newcommand{\ak}[1]{}
  \newcommand{\sd}[1]{}
	\newcommand{\ck}[1]{}
\fi

\begin{document}
\title{KernelHaven---An Open Infrastructure for Product Line Analysis}
\author{Christian Kröher, Sascha El-Sharkawy, Klaus Schmid}
\affiliation{
  \institution{University of Hildesheim, Institute of Computer Science}
  \streetaddress{Universit{\"a}tsplatz 1}
  \city{Hildesheim}
	\country{Germany}
  \postcode{31134}
}
\email{{kroeher, elscha, schmid}@sse.uni-hildesheim.de}

\begin{abstract}
KernelHaven is an open infrastructure for Software Product Line (SPL) analysis. It is intended both as a production-quality analysis tool set as well as a research support tool, e.g., to support researchers in systematically exploring research hypothesis. For flexibility and ease of experimentation KernelHaven components are plug-ins for extracting certain information from SPL artifacts and processing this information, e.g., to check the correctness and consistency of variability information or to apply metrics. A configuration-based setup along with  automatic documentation functionality allows different experiments and supports their easy reproduction.

Here, we describe KernelHaven as a product line analysis research tool and highlight its basic approach as well as its fundamental capabilities. In particular, we describe available information extraction and processing plug-ins and how to combine them. 
On this basis, researchers and interested professional users can rapidly conduct a first set of experiments. Further, we describe the concepts for extending KernelHaven by new plug-ins, which reduces development effort when realizing new experiments.
\end{abstract}

%
%

\copyrightyear{2018}
\acmYear{2018}
\setcopyright{rightsretained}
\acmConference[SPLC '18]{22nd International Systems and Software
Product Line Conference - Volume B}{September 10--14,
2018}{Gothenburg, Sweden}
\acmBooktitle{22nd International Systems and Software Product Line
Conference - Volume B (SPLC '18), September 10--14, 2018, Gothenburg,
Sweden}\acmDOI{10.1145/3236405.3236410}
\acmISBN{978-1-4503-5945-0/18/09}

\maketitle

\section{Introduction}
\label{sec:Introduction}
KernelHaven is a tool for supporting professional research, in particular, in the area of Software Product Line (SPL) analysis and verification. In order to provide optimal research support, it rests on the underlying metaphor of an \textit{experimentation workbench}: it is explicitly designed to support rapid evaluation of various research hypotheses, easy creation of different setups in the exploration stage, reuse of research results, and support of reproducibility and documentation as described in \cite{KroeherEl-SharkawySchmid18}. It thus aims to support researchers in conducting experiments in a similarly broad way as a development workbench like Eclipse supports a developer.

While KernelHaven explicitly focuses on product line analysis and verification, it aims to support a wide range of tasks in this domain and due to its openness is closer to an open ecosystem than to a product line. As we discuss in Section~\ref{sec:Overview}, it currently supports (up to) three different input pipelines, which are fed into a processing step typically addressing analysis or verification activities. 
All individual steps in extraction and processing can be realized by different, independent plug-ins, creating a broad range of possible applications. There exist already a significant number of publicly available plug-ins. Some are actually  wrapped components of well-known research tools like Undertaker~\cite{Undertaker} or TypeChef~\cite{TypeChef}, while others have been developed as part of the KernelHaven-project\footnote{\label{fn:Website}\url{https://github.com/KernelHaven}}. Due to the integration framework that KernelHaven offers, it is easy to set up an experiment combining code analysis from Undertaker with a variability analysis using KConfigReader~\cite{KConfigReader}. And it is also a matter of minutes to determine whether results differ, if as an alternative for code-analysis the Code Block extraction plug-in is used.
As a practical consideration it is sometimes also important that as a result of this clear separation into plug-ins, the various tools may have different licenses and can still be used in combination as long as the user is allowed to use each individual part.\trim
 
In its current form, especially given the currently available plug-ins, KernelHaven is well equipped for Linux-related analysis. However, we emphasize, that there is no architectural restriction in this regard. Rather, it can be used for any type of product line and any type of static analysis. The architecture is indeed so open that it can also be used itself as some form of post- or pre-processing in other scientific analysis. We are currently investigating such applications. 

Besides the support for research, the well-defined packaging of the various plug-ins allows to apply these analyses more easily and flexibly in industry. We experienced this in some of our industrial cooperations, where the framework greatly enhanced the usability also for the industrial partners \cite{El-SharkawyJyotiKrafczyk+18}.  

The remainder of this paper is structured as follows: Section~\ref{sec:Overview} provides a short overview of the basic architecture of the tool. Section~\ref{sec:Capabilities} then looks more in-depth into the current capabilities, in particular, the different available plug-ins. The intention is to provide a better understanding of the out-of-the-box capabilities of the tool as it is available right now. 
Section~\ref{sec:Extensibility} reviews the  approach for extensibility of the framework. This is particularly relevant to anyone aiming to integrate fundamentally new analysis approaches. Finally, in Section~\ref{sec:Conclusion}, we conclude.

\section{Overview}
\label{sec:Overview}
\begin{figure}[t] 
	\centering
		\includegraphics[width=\columnwidth,trim={6,8cm 7,1cm 6,5cm 5,1cm},clip]{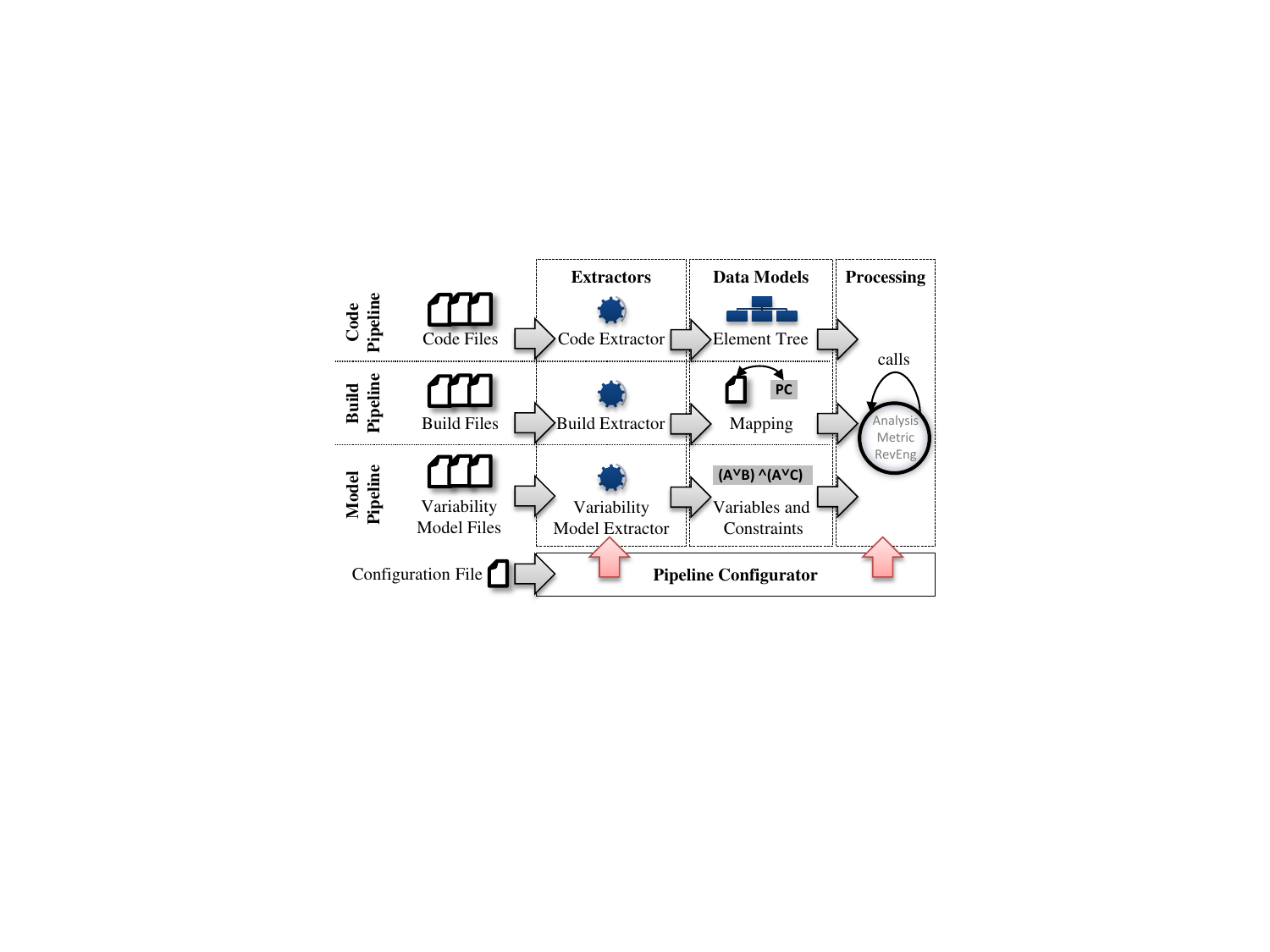}
  \caption{KernelHaven architecture}
	\label{fig:Architecture}
\end{figure}

KernelHaven is designed to support a variety of similar experiments in the domain of static SPL analysis and verification \cite{KroeherEl-SharkawySchmid18}. This design enables rapid prototyping as well as detailed analyses by offering an open plug-in infrastructure implemented in Java. The core components of this infrastructure are three extraction pipelines, the data processing, and a pipeline configurator as illustrated in Figure~\ref{fig:Architecture} and described in this section.

Each \textbf{pipeline} extracts and provides information of a particular type of artifact typically considered in variability-based analyses:

\begin{itemize}
	\item The code pipeline in the upper part of Figure~\ref{fig:Architecture} extracts information from code files resulting in a set of element trees. An element tree represents a single code file and provides information about the available code elements on different levels of abstraction. We provide more details on this data model in Section~\ref{sec:Capabilities}.
\item The build pipeline in the middle of Figure~\ref{fig:Architecture} extracts and provides information from build files. The result of this extraction is a map of files and their presence conditions (PC in Figure~\ref{fig:Architecture}) \cite{DietrichTartlerSchroderPreikschat+12, NadiHolt12}. These conditions define constraints, which must be satisfied to compile and link or, in general, build a specific (set of) file(s).
\item The (variability) model pipeline in the lower part of Figure~\ref{fig:Architecture} translates information from variability model files into a list of features and propositional formulas. They represent the features and constraints, which define the planned products of the SPL.
\end{itemize}

The quality of the extracted information depends on the specific extractors defined for the individual pipelines. Extractors can be easily exchanged by adapting the configuration file described later in this section. This exchangeability allows, e.g., to use a fast but imprecise extractor for early experiments, which serve the improvement of the experiments core processes and algorithms, and to switch to a rather slow but detailed extractor for final results.

The information provided by one or multiple pipelines is input to the core \textbf{processes and algorithms} of the conducted experiment. As experiments can be of different nature, the particular processing of the input information depends on the desired results. Hence, this processing can be exchanged similar to the extractors. We discuss the already available processing variants in Section~\ref{sec:Capabilities}, ranging from static analysis and verification tasks to software metrics. A particular feature of KernelHaven processing plug-ins is that they can call other processing plug-ins to consume their results, like the configuration mismatch analysis described in Section~\ref{sec:Capabilities}. Hence, there can be a complete sub-pipeline for processing, too.

The \textbf{configuration} file in the lower left part of Figure~\ref{fig:Architecture} defines the particular setup of an experiment and, hence, a specific instance of KernelHaven. It consists of a set of parameters for preparing the infrastructure (input and output locations, etc.) as well as defining the required pipelines, the desired extractors, and processing plug-ins. The pipeline configurator reads these parameters to configure KernelHaven prior to its execution.

Besides these core components and the resulting flexibility in setting up different experiments, KernelHaven also offers features, which particularly address common research challenges \cite{Boettiger15, CitoGall16, EichelbergerSassSchmid16}: comprehensive documentation and reproduction of experiments. KernelHaven supports automatic archiving of all relevant plug-ins, input, output, and intermediate data along with its core infrastructure. Further, the resulting archive contains the configuration file, which defines the experimental setup and, hence enables reproduction of an entire experiment on a new machine. This only requires the installation of Java on an operating system, which the extractors of the experiment support. In the following section, we explicitly include these dependencies on specific operating systems as part of our discussion of existing KernelHaven capabilities.

\section{Capabilities}
\label{sec:Capabilities}
The flexibility in combining and exchanging extractors and processing plug-ins is one of KernelHaven’s core capabilities, which, in particular, aims at supporting different users to conduct their desired analysis. As a starting point, different plug-ins are publicly available, which can be used out of the box. In this section, we introduce them by, first, discussing available extractors, their purpose, and extracted information and, second, describing realized processing plug-ins and their possible combinations with the extractors.\trim

\begin{figure}[t]
	\centering
		\includegraphics[trim={3.5cm 9.35cm 6.3cm 2.8cm},clip,width=\columnwidth]{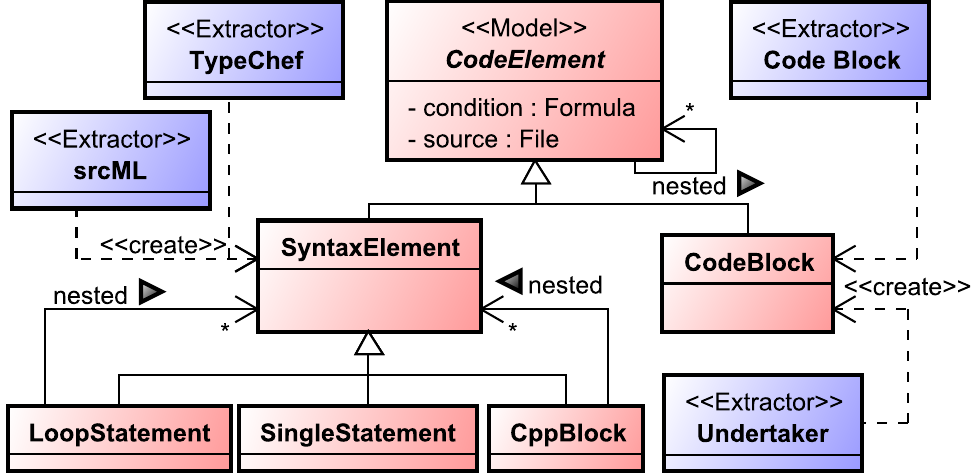}
	\caption{Simplified code model of KernelHaven}
	\label{fig:Code Model}
\end{figure}

KernelHaven provides seven different extractors as summarized in Table~\ref{tab:extractors}. Most of these extractors are \textbf{code extractors}. They rely on the code model illustrated in a simplified form in Figure~\ref{fig:Code Model} to provide code information of different granularity: ranging from code blocks to entire Abstract Syntax Trees (ASTs). This range of expressiveness is required to support a variety of external tools wrapped as KernelHaven plug-ins, like \textbf{Undertaker} \cite{Undertaker}, \textbf{srcML} \cite{srcML}, and \textbf{TypeChef} \cite{TypeChef}. Except for Undertaker, which is typically available as native Linux binary, these extractors are executable on any operating system. Due to this limitation of Undertaker, we also provide a reimplementation of its algorithms in Java resulting in the operating-system-independent and faster \textbf{Code Block} extractor. Hence, both extractors extract conditional code blocks (right side of Figure~\ref{fig:Code Model}) from C-code files defined as input in Table~\ref{tab:extractors}. In contrast, srcML and TypeChef provide ASTs of their input files, which results in a set of possibly nested \texttt{SyntaxElement}s as illustrated on the left side of Figure~\ref{fig:Code Model}. While the AST of the srcML extractor currently only provides a granularity of single line statements\footnote{This is due to the ongoing work on this extractor; the original srcML tool provides more details.}, the TypeChef extractor enables variability-aware parsing including macro-expansion\footnote{TypeChef automatically considers header files (\texttt{*.h}) without an explicit definition of those files as input.}, which results in significantly more details about code files.\trim

\begin{table}
	\caption{Available KernelHaven extractors (April 2018)}
	\begin{threeparttable}
	\centering
		\begin{tabulary}{\columnwidth}{LLLLL}
			\hline
			\textbf{Name} & \textbf{Type} & \textbf{Input} & \textbf{OS} \\ \hline \hline
			Undertaker\tnote{1} & Code & *.c/*.h/*.S & Linux \\ \hline
			Code Block\tnote{2} & Code & *.c/*.h/*.S & Any \\ \hline
			srcML\tnote{1} & Code & *.c/*.h & Any \\ \hline
			TypeChef\tnote{1} & Code & *.c & Any \\ \hline
			KBuildMiner\tnote{1} & Build & KBuild/Make & Any \\ \hline
			KConfigReader\tnote{1} & VM & Kconfig & Linux \\ \hline
			DIMACSVariabilityModel\tnote{2} & VM & DIMACS & Any \\ \hline
		\end{tabulary}
		\begin{tablenotes}
			\footnotesize
			\item VM = Variability Model 
			\item[1] This extractor wraps an external tool not developed by the authors
			\item[2] This extractor is solely developed by the authors
		\end{tablenotes}
	\end{threeparttable}
	\label{tab:extractors}
\end{table}

The \textbf{KBuildMiner} extractor is the only publicly available \textbf{build extractor} so far. It wraps the corresponding tool \cite{KBuildMiner}, which takes KBuild- and Make-files as input and provides the mapping of files and presence conditions as described in Section~\ref{sec:Overview}. In particular, KBuild-files stem from the Linux kernel build system \cite{Kbuild18} and its specific variability realization technique, which uses \texttt{CONFIG\_*} symbols in code and build files to reference configuration options defined in the variability model. Hence, this extractor explicitly searches for such references to build the mapping. Build extractors for other variability realization techniques can easily be implemented by exploiting the extensibility features described in Section~\ref{sec:Extensibility}.\trim

The \textbf{KConfigReader} and the \textbf{DIMACSVariabilityModel} extractors represent two fundamentally different \textbf{variability model extractors}. While the former wraps the KConfigReader tool \cite{KConfigReader}, which explicitly takes Kconfig-files \cite{Kbuild18} as input and uses build tools of the Linux kernel restricting its applicability to Linux only, the latter extracts propositional formulas in DIMACS format and converts them into the internal variability model representation of KernelHaven on any operating system. Similar to the KBuildMiner build extractor, KConfigReader is the extractor of choice if an experiment requires variability model information from the Linux kernel build system. The DIMACSVariabilityModel extractor can be used, for example, for any SPL using a feature model as this type of model can be easily translated into propositional formulas \cite{Batory05} and, hence, into DIMACS format.

KernelHaven allows any combination of the above extractors for a specific experiment. However, the actual processes and algorithms implemented in a processing plug-in may require a particular setup of the pipelines. Table~\ref{tab:processings} includes such dependencies for the available processing plug-ins as an example. The \textbf{UnDeadAnalysis} detects dead code blocks or missing features as described in \cite{TartlerLohmannSincero+11}. Therefore, it requires the respective block information from code files, which multiple code extractors provide. Hence, one of them needs to be defined in the respective KernelHaven configuration file. Further, the analysis requires build information to determine whether the code file including the respective code block is part of the final product. The availability of a particular file and its inherent code blocks is typically controlled by configuration options defined in the variability model, which requires a corresponding extractor.

\begin{table*}[!bt]
	\caption{Available KernelHaven processing plug-ins and possible extractor combinations (July 2018)}
	\vspace*{-1em}
	\begin{threeparttable}
	\centering
		\begin{tabular}{lp{5cm}p{3cm}ll}
			\hline
			\textbf{Name} & \textbf{Goal} & \textbf{Code Extractor} & \textbf{Build Extractor} & \textbf{VM Extractor} \\ \hline \hline
			UnDeadAnalysis\textsuperscript{1} & Detection of dead code blocks (never part of any product) or missing features & Undertaker, Code Block, srcML, or TypeChef & KBuildMiner & KConfigReader \\ \hline
			ConfigurationMismatches\textsuperscript{2} & Detection of inconsistencies between modeled and effective variability & \multicolumn{3}{p{8cm}}{\centering\textit{Processes the results of the FeatureEffect analysis and, hence, \newline does not directly use a particular extractor}} \\ \hline
			FeatureEffect\textsuperscript{1} & Reverse engineering of variability constraints from code and build files to improve an existing or create a new VM & Undertaker, Code Block, srcML, or TypeChef & KBuildMiner & \textit{not required}\\ \hline
			MetricHaven\textsuperscript{1,2} & More than 7,300 metrics and their variants for measuring SPL properties & srcML (TypeChef possible, but not realized yet) & KBuildMiner & KConfigReader \\ \hline
		\end{tabular}
		\begin{tablenotes}
			\footnotesize
			\item VM = Variability Model; \textsuperscript{1} This processing is a reimplementation in KernelHaven of existing work; \textsuperscript{2} This processing is developed by the authors
		\end{tablenotes}
	\end{threeparttable}
	\label{tab:processings}
	\vspace*{-7pt}
\end{table*}

The \textbf{FeatureEffect} analysis requires a different setup of KernelHaven as it is a reverse engineering approach, which identifies variability constraints from code and build files \cite{NadiBergerKastner+15}. Hence, it does not need a variability model extractor as well as the respective pipeline. For this purpose, a user simply excludes the corresponding parameters from KernelHaven’s configuration file. Further, the \textbf{ConfigurationMismatches} analysis builds upon the same configuration as it uses the results of the feature effect analysis to check whether the implemented (effective) variability matches the variability model \cite{El-SharkawyKrafczykSchmid17}. This concatenation of processing plug-ins results in processing or analysis pipelines, which we describe in detail in \cite{KroeherEl-SharkawySchmid18}. In essence, processing pipelines are defined by combining processing plug-ins in the order they consume their results in the configuration file without any further implementation effort.

A fundamentally different type of experiment is represented by \textbf{MetricHaven}. This processing plug-in
provides currently more than 7,300 metric variations \cite{El-SharkawyYamagishi-EichlerSchmid17} to measure SPL artifact types in isolation as well as in combination. Similar to the \textbf{UnDeadAnalysis}, this plug-in requires all pipelines and, in particular, srcML as it provides the necessary implementation details.

The set of already available plug-ins shows the potential of KernelHaven to support various types of experiments ranging from verification and analyses to metrics. A user can conduct these experiments with different levels of details, e.g., code blocks or entire ASTs, depending on the configured extractors. Further, new plug-ins can be realized and combined with the existing ones to extend the current capabilities as described in the following section.

\section{Extensibility}
\label{sec:Extensibility}
The architecture of KernelHaven is designed to simplify the development of new extensions by means of separation of concerns and reuse. This allows researchers to focus on the realization of novel concepts while abstracting from technical details. We outline the three design concepts for achieving this simplicity in this section.

\textbf{Decoupled extractors and processing.} Independent data models decouple the extraction of relevant information and the processing of the extracted data. On the one side, this facilitates the reuse of already developed processing plug-ins for new artifact types as only a new extractor is required. One the other side, the realization of novel processing algorithms, e.g., a new code metric, may reuse existing extractors for already handled artifact types.

\textbf{Reuse of processing algorithms.} KernelHaven processing plug-ins do not differentiate between the output of extractors or other processing plug-ins. Thus, new processing plug-ins may reuse existing and mature processing plug-ins to form a processing or analysis pipeline, like the concatenation of the feature effect and configuration mismatches analyses in Section~\ref{sec:Capabilities}. In this way, developers can realize complex processing algorithms incrementally. Individual KernelHaven plug-ins may represent these increments, which can be used both in isolation as well as in combination with other processing algorithms and, hence, avoid extra effort for re-implementing required processing capabilities. 

\textbf{Abstraction from technical details.} The KernelHaven infrastructure separates conceptual work from technical implementation. This way, researchers do not need to consider aspects like multithreading or saving of results as the infrastructure takes care of that. This means that KernelHaven executes by default all extraction pipelines (cf.\ Figure~\ref{fig:Architecture}) and all processing plug-ins in parallel. For instance, the configuration mismatch analysis may start to validate that the variability model covers the detected code dependencies of the feature effect analysis, while this analysis still detects further code dependencies. However, developers can force all processing components (extractors or processing plug-ins) to complete before a new component starts. To save the results, a developer only needs to annotate data classes and KernelHaven takes care of storing the results. Thus, developers usually neither need to specify any output formats nor to take care of stream handling.

\vspace*{-1ex}
\section{Conclusion}
\label{sec:Conclusion}
We presented KernelHaven as an experimentation workbench to support different types of users in their goal to conduct their individual experiments in the domain of static Software Product Line (SPL) analyses and verification.  Users benefit from the available plug-ins, which wrap some of the most prominent analysis tools in this domain and, hence, simplify their application even in new experiments. KernelHaven addresses researchers by explicitly supporting the documentation and reproduction of experiments. Developers take advantage of the rather simple extensibility due to reusability of existing plug-ins and abstraction from technical details provided by the core infrastructure.
We believe that KernelHaven is also applicable in teaching SPL engineering as it significantly reduces time and effort students need to setup a particular extraction, analysis, metric, etc. Rather, it is simply a matter of extracting an archive and executing KernelHaven with the desired configuration.

\vspace*{-1ex}
\begin{acks}
This work is partially supported by the ITEA3 project $\text{REVaMP}^2$, funded by the \grantsponsor{01IS16042H}{BMBF (German Ministry of Research and Education)}{https://www.bmbf.de/} under grant \grantnum{01IS16042H}{01IS16042H}. Any opinions expressed herein are solely by the authors and not by the BMBF.
  
We thank the following contributors: Moritz Fl\"oter, Adam Krafczyk, Alice Schwarz, Kevin Stahr, Johannes Ude, Manuel Nedde, Malek Boukhari, and Marvin Forstreuter.
\end{acks}

\begin{spacing}{0.98}
  \bibliographystyle{ACM-Reference-Format}
  \bibliography{literature}
\end{spacing}

\appendix
\section{Demonstration and Case Studies}
\label{sec:Appendix}
\begin{figure}[ht] 
	\centering
		\includegraphics[width=0.9\columnwidth,trim={0cm 9,8cm 13,4cm 0cm},clip]{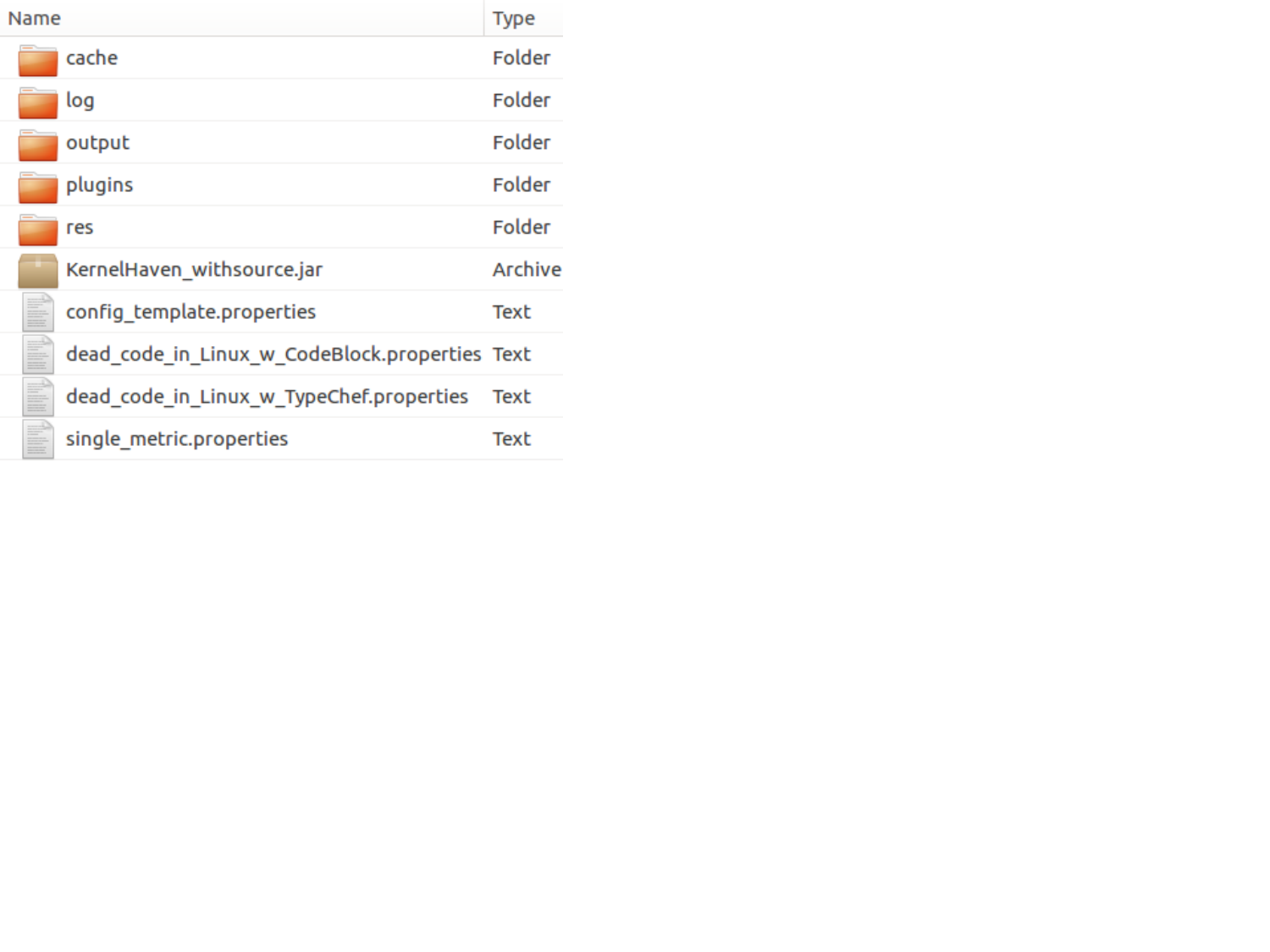}
  \caption{KernelHaven bundle content (example)}
	\label{fig:BundleContent}
\end{figure}

The KernelHaven demonstration starts with downloading a bundled release from the main infrastructure website\footnote{\url{https://github.com/KernelHaven/KernelHaven\#downloads}}. A bundle contains everything needed to conduct available analyses out of the box. Figure~\ref{fig:BundleContent} shows an example of the extracted content of such a bundle, which we briefly describe to understand the default structure of the tool and the purpose of the included files and directories.

In Figure~\ref{fig:BundleContent} the bundle contains the main infrastructure (\texttt{KernelHaven\_withsource.jar}), the extraction and processing plug-ins (\texttt{plugins} directory), and a set of configuration files (\texttt{*.properties}) defining a particular setup of the tool. These configurations use the included set of empty directories, e.g., for saving cached information (\texttt{cache}), log files (\texttt{log}), and the final analysis results (\texttt{output}). Hence, we can start the tool, for example to collect metrics, on a command line prompt\footnote{KernelHaven is a pure command line tool; it does not provide a graphical user interface} by calling the main infrastructure with the metrics configuration file as single parameter:

\Command{\texttt{java~-jar~KernelHaven\_withsources.jar \newline \indent\indent \mbox{\textcolor{red}{$\hookrightarrow$}} single\_metric.properties}}

Listing~\ref{lst:Config} shows an excerpt of the \texttt{single\_metric.properties} file, which contains a set of information blocks that we explain next. The \textbf{directories} in Lines~\ref{line:directories_start}\,--\,\ref{line:directories_end} specify the mandatory directories, like the location of the software product line to analyze (\texttt{source\_tree}), the location to store the results of the analysis (\texttt{output\_dir}), or the location of the extractors and analyses (\texttt{plugins\_dir}). While a bundle already contains those directories, the respective parameters can be used to specify custom ones. Lines~\ref{line:logging_start}\,--\,\ref{line:logging_end} specify the amount of \textbf{logging} information (\texttt{log.level}), which the infrastructure prints to a file (\texttt{log.file}) in the specified directory (\texttt{log.dir}) and, optionally, to the console (\texttt{log.console}). The \textbf{code model parameters} in Lines~\ref{line:code_start}\,--\,\ref{line:code_end} define the 
\begin{figure}[ht]
	\centering
		\vspace{5pt}
\begin{lstlisting}[caption={\texttt{single\_metric.properties} file (excerpt), place holder for local settings are marked as \color{cEclipsePurple}\texttt{< >}\color{black}.},label=lst:Config,basicstyle=\small,language=Properties]
###  Directories   ###
resource_dir =	res/               %*\label{line:directories_start}*)
output_dir =	output/
plugins_dir =	plugins/
cache_dir =	cache/
source_tree = <path/to/SPL>      %*\label{line:directories_end}*)

###  Logging  ###
log.dir =	log/                   %*\label{line:logging_start}*)
log.console =	false
log.file =	true
log.level =	INFO                   %*\label{line:logging_end}*)

###   Code Model Parameters  ###
code.provider.timeout =	0          %*\label{line:code_start}*)
code.provider.cache.write =	false
code.provider.cache.read =	false
code.extractor.class =	net.ssehub.kernel_haven.srcml.SrcMLExtractor
code.extractor.file_regex =	.*\.c  %*\label{line:code_end}*)

###  Build Model Parameters  ###
build.provider.timeout =	0      %*\label{line:build_start}*)
build.provider.cache.write =	true
build.provider.cache.read =	true
build.extractor.class =	net.ssehub.kernel_haven.kbuildminer.KbuildMinerExtractor  %*\label{line:build_end}*)

###   Variability Model Parameters  ###
variability.provider.timeout =	0   %*\label{line:varModel_start}*)
variability.provider.cache.write =	true
variability.provider.cache.read =	true
variability.extractor.class =	net.ssehub.kernel_haven.kconfigreader.KconfigReaderExtractor  %*\label{line:varModel_end}*)

###   Analysis Parameters  ###
analysis.class =	net.ssehub.kernel_haven.metric_haven.metric_components.MetricsRunner %*\label{line:analysis_start}*)
analysis.metrics_runner.metrics_class = <class name of metric to execute>
metrics.round_results =	true
analysis.output.type =	xlsx        %*\label{line:analysis_end}*)
\end{lstlisting}
		\vspace*{-1em}
\end{figure}
availability of the code pipeline in general and, in particular, the code extractor (\texttt{code.extractor.class}) via its fully qualified main class name. Further, the parameters define extracting information from all files in the \texttt{source\_tree} matching a given regular expression (\texttt{code.extractor.file\_regex}), which only identifies C-code class files. While a specific timeout for automatically terminating the extractor (\texttt{code.provider.timeout}) is not defined, caching intermediate code data (\texttt{code.provider.cache.write} and \texttt{code.provider.cache.read}) is disabled as it is currently not supported by the \texttt{SrcMLExtractor}. Similar to the code model parameters, Lines~\ref{line:build_start}\,--\,\ref{line:build_end} define the \textbf{build model parameters}, like the build extractor (\texttt{build.extractor.class}), the timeout and the caching capabilities. The \textbf{variability model parameters} in Lines~\ref{line:varModel_start}\,--\,\ref{line:varModel_end} specify the variability model extractor (\texttt{variability.extractor.class}) for the corresponding pipeline. Similar to the previous pipeline no timeout but caching is enabled. In order to exclude a pipeline from an experiment, only the respective block of parameters has to be deleted from the configuration.

The last block of information in Lines~\ref{line:analysis_start}\,--\,\ref{line:analysis_end} of Listing~\ref{lst:Config} specifies the actual processing algorithm to use for this particular experiment by the fully qualified class name (\texttt{analysis.class}). In this example, the \texttt{MetricsRunner} is used, to load a metrics analysis and to execute all variations of the specified metric. This processing algorithm requires an additional parameter \texttt{ana\-ly\-sis.me\-trics\_runner.metrics\_class} to select the particular metric algorithm to be executed. In the example, we choose \texttt{net.sse\-hub.kernel\_haven.metric\_haven.metric\_componen\-ts.DLoC} to compute three variations of  the \textit{Delivered Lines of Code}-metric for each code function of the analyzed product line. The last two parameters limit the output to two decimal places and specify that the results should be saved as a CSV-file. The KernelHaven bundles are shipped with a full documentation of available configuration settings (\texttt{config\_template.properties}).

\begin{table}
	\centering
		\caption{DLoC-metric results for Linux 4.15 (excerpt).}
		\vspace*{-1em}
		\begin{tabular}{|l|l|l|l|l|l|}
		\hline
		\textbf{Source}&\textbf{Line}&\textbf{Function}&\textbf{DLoC}&\textbf{LoF}&\textbf{PLoF}\\
		\hline
		\begin{filecontents*}{figures/DLoC.csv}
Source File;Line No.;Element;LoC;LoF;PLoF
a20.c;100;enable\_a20\_kbc;7.0;0.0;0.0
build.c;248;efi\_stub\_defaults;3.0;3.0;1.0
build.c;282;efi\_stub\_defaults;0.0;0.0;0.0
cpu.c;37;show\_cap\_strs;21.0;20.0;0.95
cpu.c;75;validate\_cpu;16.0;0.0;0.0
eboot.c;748;setup\_e820;63.0;1.0;0.01
eboot.c;877;exit\_boot\_func;23.0;2.0;0.08
eboot.c;964;efi\_main;116.0;0.0;0.0
misc.c;183;handle\_relocations;33.0;14.0;0.42
\end{filecontents*}
\csvreader[separator=semicolon, head to column names=true, late after line=\\\hline]
			{figures/DLoC.csv}
			{}
			{\csvcoli & \csvcolii & \csvcoliii & \csvcoliv & \csvcolv & \csvcolvi}
	\end{tabular}
	\label{tab:DLoC Result}
\end{table}

Table~\ref{tab:DLoC Result} shows an excerpt of the analysis results for the three variations of the DLoC-metric. The first three columns list the measured item: path of the measured C-file (folder names were removed for the sake of brevity), the line number, and the name of the measured function. The last three columns show the measuring results for the DLoC, the Lines of Feature code (LoF), and the fraction of feature code ($\text{PLoF} = \dfrac{\text{LoF}}{\text{DLoC}}$) \cite{El-SharkawyYamagishi-EichlerSchmid17}. The results show two separate implementations of the function \texttt{efi\_stub\_defaults} at the Lines~248 and 282 of the file \texttt{build.c}. The first implementation contains three statements completely surrounded by inner \ifDef-blocks (DLoC and LoF have the same value and, thus, the PLoF value is 100\%). The second implementation is an empty function definition for the case that the feature ``EFI\_STUB'' was deselected.\trim

Another supported analysis is the detection of code dependencies, which was also applied at the Bosch PS-EC product line \cite{El-SharkawyJyotiKrafczyk+18}. For confidentiality reasons, we cannot show details specific for Bosch, but the general part of the analysis is publicly available and may also be applied to open source system like Linux. However, this analysis requires a different setup. The most important changes are presented in Listing~\ref{lst:Config-FE}. In Line~\ref{line:code_extractor} the code extractor is exchanged to a faster and more lightweight code block extractor. This extractor does not extract C-code statements, which are not required for the analysis. The extractor considers only the relevant C-preprocessor statements and their conditions. Further, it provides support to handle macro definitions, which is necessary to handle the variability information of Linux correctly (cf.\ Line~\ref{line:code_macros}). Line~\ref{line:code_fails} specifies that unparsable conditions shall be treated as \texttt{true}. This allows the extractor to continue parsing a file if a condition cannot be handled rather than skipping the complete file.

\begin{figure}[ht]
	\centering
\begin{lstlisting}[caption={Modified properties file to analyze code dependencies (delta).},label=lst:Config-FE,basicstyle=\small,language=Properties]
###   Code Model Parameters  ###
code.extractor.class =	block_extractor.CodeBlockExtractor  %*\label{line:code_extractor}*)
code.extractor.handle_linux_macros =	true  %*\label{line:code_macros}*)
code.extractor.invalid_condition =	true      %*\label{line:code_fails}*)

###   Analysis Parameters  ###
analysis.class =	analysis.ConfiguredPipelineAnalysis %*\label{line:analysis}*)
analysis.pipeline =	                     \ %*\label{line:analysis_dsl_start}*)
	 fe_analysis.fes.FeatureEffectFinder(   \
	   fe_analysis.pcs.PcFinder(            \
	     cmComponent(), bmComponent()       \
	   )                                    \
	 ) %*\label{line:analysis_dsl_end}*)
analysis.output.type =	xlsx %*\label{line:analysis_Excel}*)
analysis.output.intermediate_results =	PcFinder%*\label{line:analysis_intermediate}*)
analysis.consider_vm_vars_only =	true%*\label{line:analysis_filter}*)
\end{lstlisting}
		\vspace*{-1em}
\end{figure}

Lines~\ref{line:analysis}\,--\,\ref{line:analysis_filter} specify the analysis. First, \texttt{net.ssehub.kernel\-\_ha\-ven.ana\-lysis.ConfiguredPipeline\-Ana\-lysis} specifies a wiring of a (complex) processing pipeline via an embedded DSL (cf.\ Lines~\ref{line:analysis_dsl_start}\,--\,\ref{line:analysis_dsl_end}). This pipeline consists of two processing algorithms. The first algorithm gathers for each feature all presence conditions in which it is used in (\texttt{PcFinder}). The second algorithm combines all presence conditions to compute a condition under which the selection of a feature has an effect on the product derivation (\texttt{FeatureEffectFinder}). By default, KernelHaven stores only the results of the last processing algorithm. Lines~\ref{line:analysis_Excel} and \ref{line:analysis_intermediate} specify that both results should be saved into separate sheets of an Excel workbook (*.xlsx). The last line is used to limit the results only to features specified in the variability model, otherwise all constants used in C-preprocessor statements will be collected. \texttt{cmComponent}, \texttt{bmComponent}, and \texttt{vmComponent} (not used in Listing~\ref{lst:Config-FE}) are pre-defined elements of KernelHaven's DSL to specify that a processing algorithm shall take the information of the code, build, or variability model extractor as input. Please note that we removed the prefix \texttt{net.ssehub.kernel\-\_ha\-ven} from all class names in Listing~\ref{lst:Config-FE} to improve readability.

Table~\ref{tab:FE Result} presents an excerpt of the results of the feature effect analysis. The first column lists all features found in code files or in Make-files. The second column shows the precondition, which must be fulfilled in order that the selection of the feature has an effect of the product derivation. For instance, the selection of \texttt{CONFIG\_ACPI\_DEBUG} and \texttt{CONFIG\_ACPI\_DEBUGGER} change the code only if also \texttt{CONFIG\_ACPI} is selected. This may happen if both features are always nested below \texttt{CONFIG\_ACPI}, while \texttt{CONFIG\_ACPI} may also be used as top-level statement.
\begin{table}[bh]
	\centering
		\vspace*{-5pt}
		\caption{Feature Effect results for Linux 4.15 (excerpt).}
		\vspace*{-1em}
		\begin{tabular}{|l|p{3.5cm}|}
		\hline
		\textbf{Feature}&\textbf{Relevancy Condition}\\
		\hline
		\begin{filecontents*}{figures/FEs.csv}
Feature;Precondition
CONFIG\_ACPI;TRUE
CONFIG\_ACPI\_DEBUG;CONFIG\_ACPI
CONFIG\_ACPI\_DEBUGGER;CONFIG\_ACPI
CONFIG\_ACPI\_I2C\_OPREGION;CONFIG\_ACPI $\land$ (CONFIG\_I2C\_MODULE $\lor$ CONFIG\_I2C)
\end{filecontents*}
\csvreader[separator=semicolon, head to column names=true, late after line=\\\hline]
			{figures/FEs.csv}
			{}
			{\csvcoli & \csvcolii}
	\end{tabular}
	\label{tab:FE Result}
\end{table}

\end{document}